\documentclass[a4,preprint,superscriptaddress,floatfix, showpacs]{revtex4}
\usepackage{amsmath}
\usepackage{graphicx}
\begin{document}
\author{A.A. Avetisyan}
\email{artakav@ysu.am} %
\affiliation{Department of Physics, Yerevan State University, 1 A.
Manoogian, 0025 Yerevan, Armenia}
\author{A. P. Djotyan}
\email{adjotyan@ysu.am} %
\affiliation{Department of Physics, Yerevan State University, 1 A.
Manoogian, 0025 Yerevan, Armenia}
\author{K. Moulopoulos}
\email{cos@ucy.ac.cy} %
\affiliation{Department of Physics, University of Cyprus, P.O. Box
20537, 1678 Nicosia, Cyprus}
\date{\today}
\title{NONLINEAR PROPERTIES OF GATED GRAPHENE IN A STRONG ELECTROMAGNETIC FIELD}
\begin{abstract}
 We develop a microscopic theory of a strong electromagnetic field interaction with gated bilayer graphene.
 Quantum kinetic equations for density matrix are obtained using a tight binding approach within second quantized
 Hamiltonian in an intense laser field. We show that adiabatically changing the gate potentials with time may produce
 (at resonant photon energy) a full inversion of the electron population with high density between valence and conduction
 bands. In the linear regime,
 excitonic absorption of an electromagnetic radiation in a graphene monolayer with opened energy gap is also studied.

\end{abstract}


\maketitle

\section{Introduction}

Graphene, a two-dimensional (2D) crystal of carbon atoms packed in
hexagonal lattice, has attracted great interest last years due to
its exotic electronic properties~\cite{novoselov, castrorev}. The
charge carriers in graphene are massless Dirac fermions with
effective "velocity of light" $v_{F}=10^{6} m/s$, that by two orders
is smaller than the speed of light in vacuum.

The deep connection between the electronic properties of graphene
and certain theories in particle physics makes graphene a testbed
for many ideas in basic science. Besides the well-known electronic
properties, such as ballistic electron transport~\cite{novoselov,
castrorev, novoselovn} quantum Hall effect~\cite{zhang} tunable band
gap~\cite{rotenberger, zhang1} graphene also shows very interesting
optical properties~\cite{li}. Ultrathin graphite films with
excellent mechanical quality and fascinating energy spectrum are
very promissing, e.g. for nanoelectronics~\cite{geimrev} and as
transparent conducting layers~\cite{geimnano} which are important,
e.g. for displays and solar cells.

Graphene has been extensively considered as a promising material for
nonlinear optical applications~\cite{ishikaw}. In particular, the
nonlinear quantum electrodynamics effects can be observed in
graphene already for fields available in the
laboratory~\cite{katsnels}. In Ref.~\onlinecite{ishikaw} the
nonlinear optical response of electron dynamics in monolayer
graphene to an intense light pulse was investigated.
 The study of the nonlinear electromagnetic effects in graphene has so far mainly
focused on monolayer graphene. Meanwhile, there is growing interest
in bilayer and trilayer graphene systems, where the electronic band
structures are richer than in monolayer and can be easily
manipulated by the application of external fields. Theoretical and
experimental investigations have shown that a perpendicular electric
field (created by gates) applied to a graphene bilayer modifies its
band structure near the K point and may open an energy
gap~\cite{castro} in the electronic spectrum.
Magnetotransport~\cite{castro} and spectroscopic~\cite{zhang1}
measurements showed that the induced gap between the conduction and
valence bands could be tuned between zero and midinfrared energies.
This makes bilayer graphene the only known semiconductor with a
tunable energy gap, and may open the way for developing
photodetectors and lasers tunable by an electric field.

The magnitude of the gap also strongly depends on the number of
graphene layers and its stacking order~\cite{avetis, avetis1, bao}.
A desirable band structure for multilayer graphene systems, that can
be useful for different purposes of nano- and optoelectronics, can
be theoretically found (and suggested for an experimental
realization) e.g. by a corresponding choice of the layer number in a
graphene system. Multilayer graphene systems exhibit rich novel
phenomena at low charge densities owing to enhanced electronic
interactions~\cite{bao}. In~\cite{avetis, avetis1} we studied
theoretically electric field induced band gap of graphene
multilayers with different ways of stacking between consecutive
graphene planes. Using a positively charged top and a negatively
charged back gate it is possible to control independently the
density of electrons on the graphene layers (correspondingly the
band gap) and the Fermi energy of multilayer graphene systems.

Optical measurement techniques are widely adopted to experimentally
investigate electronic properties of graphene multilayers. To
determine the number of layers, as well as the stacking structure,
Raman spectroscopy was used in~\cite{ferr, russo}. Infrared
spectroscopy is also applied to probe the low-energy band structure.
The dependence of the infrared optical absorption on the stacking
sequence has been observed in recent experiments~\cite{mak}.

It is therefore of high relevance and of great interest to
theoretically study nonlinear optical properties of multilayer
graphene  systems with opened energy gap between valence and
conduction  bands.

In the present work we develop a microscopic theory of a strong
electromagnetic field interaction with bilayer graphene with an
energy gap opened by external gates. We study the nonlinear response
of bilayer graphene, when one-photon interband excitation regime is
induced by an intense coherent radiation. We show that at resonant
photon energy, close to the energy gap, and adiabatically changing
the gate potentials with time, one can produce full inversion of the
electron population with high density between the valence and
conduction bands.

 The coherent optical response of multilayer
graphene systems to an intense laser radiation field may reveal many
particle correlation effects. Excitons are expected to modify
strongly the optical response. In monolayer graphene, since there is
no energy gap, the Coulomb problem has no true bound states, but
resonances~\cite{peres}. A signature of the presence of the
excitonic resonances was observed in its optical properties
~\cite{kin}. The optical response of graphene with an opened energy
gap between the conduction and valence bands is dominated by bound
excitons~\cite{park}.

In this work, we also study the excitonic absorption of a monolayer
graphene with an opened energy gap. A substantial band gap in
monolayer graphene can be induced in several ways, e.g., by coupling
to substrates, electrical biasing, or
nanostructuring~\cite{castrorev}.

The plan of the paper is as follows. Section \ref{sec2} presents an
outline of the underlying theory and describes the laser interaction
with a gated graphene system. The evolution equations for the
single-particle density matrix in the presence of Coulomb
interaction are obtained in Section \ref{sec3}. Excitonic absorption
in monolayer graphene is investigated in Section \ref{sec4}. Section
\ref{sec5} contains our numerical results together with a discussion
of essential physical points. Section \ref{sec6} presents our main
conclusions.

\section{LASER INTERACTION WITH GATED BILAYER GRAPHENE } \label{sec2}

We consider here the interaction of a strong electromagnetic wave
with bilayer graphene. A perpendicular electric field created by top
and back gates~\cite{avetis} opens an energy gap in multilayer
graphene. We assume that the laser pulse propagates in the
perpendicular direction to graphene plane (XY) and the electric
field $\mathbf{E}(t)$ of pulse lies in the graphene plane.

The Hamiltonian in the second quantization formalism has the form

\begin{eqnarray}
\hat{H} &=&\int \hat{\Psi}^{+}H_{s}\hat{\Psi}d\mathbf{r}
\label{hamsec}
\end{eqnarray}

 The Hamiltonian $H_{s}$ for
bilayer, in the vicinity of the $K$ point, for energies
$\varepsilon<<\gamma_{1}$ (with $\gamma_{1}=377meV$  being the
vertical interlayer hopping parameter) can be written as (here we
omit the real spin number ):

\begin{equation}
\widehat{H}_{s}=\widehat{H}_{0}+\widehat{H}_{d},  \label{hamsin}
\end{equation}

\begin{equation}
\widehat{H}_{0}=\left(
\begin{array}{cc}
-\frac{U}{2} & -\frac{\hbar^{2}}{2m}\left( {k}_{x}-i{k}_{y}\right)
^{2}+\mathrm{v}_{3}\hbar\left( {k}_{x}+i{k}_{y}\right) \\
-\frac{\hbar^{2}}{2m}\left( {k}_{x}+i{k}_{y}\right) ^{2}+\mathrm{v}%
_{3}\hbar\left( {k}_{x}-i{k}_{y}\right) & \frac{U}{2}%
\end{array}
\right),
\end{equation}

\begin{equation}
\widehat{H}_{d}=\left(
\begin{array}{cc}
e\mathbf{r \cdot E}\left( t\right) & 0 \\
0 & e\mathbf{r \cdot E}\left( t\right)%
\end{array}%
\right).
\end{equation}
The first term in Eq.~(\ref{hamsin}) corresponds to bilayer graphene
in the field of perpendicular electric field: $U$ is the gap
introduced by the perpendicular electric field, and the second term
is the interaction Hamiltonian between a laser field and bilayer
graphene; $\mathrm{v}_{3}\approx \mathrm{v}_{F}/8$ is the effective
velocity $\mathrm{v}_{3}=\sqrt{3}\gamma_3 a/2\hbar$ where
$\gamma_{3}$ describes the interaction between $B$ atoms in the
neighbouring layers, and $a$ is the lattice constant. The account of
this interaction leads to the so called triagonal warping effect of
the bands~\cite{avetis}. The Fermi velocity
$\mathrm{v}_{F}=\sqrt{3}\gamma_0 a/2\hbar\simeq10^6m/s$ and
tight-binding parameter $\gamma_{0}$ describes the interaction
between $A$ and $B$ atoms in the same layer,
$\textbf{p}=\hbar\textbf{k } $ is the electron momentum and
$\textbf{k}=\{k_{x},k_{y}\}$.

We expand the fermionic field operator over the free wave function
$\psi _{\sigma }(\mathbf{k})$ of bilayer graphene
\begin{equation}
{\Psi}(\mathbf{r},t)=\sum_{\sigma \mathbf{k}}\widehat{a}_{\mathbf{k}%
,\sigma }(t)\Psi _{\sigma }(\mathbf{k}) e^{%
i\mathbf{kr}} \\ \label{f_oper}
\end{equation}%
where the annihilation operator $\widehat{a}_{\mathbf{p},\sigma}(t)$
is associated with positive and negative energy solutions
$\sigma=\pm1$. Introducing $\vartheta \left( \mathbf{k}\right)
=\arctan ({k_{y}}/{ k_{x}}), \qquad k_{x}+ik_{y}=ke^{i\vartheta
}\qquad $ the expression for energy spectrum of bilayer graphene can
be brought to the form
\begin{eqnarray}
\mathcal{E}_{ \mathbf{k}\sigma } &=&\sigma \sqrt{\frac{U^{2}}{4}+(
\mathrm{v}_{3}\hbar k)^{2}- \frac{\mathrm{v}_{3}\hbar^{3}
k^{3}}{m}\cos 3\vartheta +(\frac{\hbar^{2}k^{2}}{2m})^{2}}
\label{energ}
\end{eqnarray}
The free solutions in bilayer graphene $\psi _{\sigma }(\mathbf{k})$
have the following form
\begin{equation}
\psi _{\sigma }(\mathbf{k}) =\sqrt{\frac{\mathcal{E}_{ \mathbf{k}
\sigma }+U/2}{2\mathcal{E}_{ \mathbf{k}\sigma }}}\left(
\begin{array}{c}
1 \\
\frac{1}{\mathcal{E}_{ \mathbf{k}\sigma
}+U/2}\Upsilon\left(k,\vartheta\right)
\end{array}
\right), \label{spinor}
\end{equation}
where
\begin{equation}
\Upsilon ( k,\vartheta ) =-\frac{\hbar^{2} k^{2}}{2m}e^{i2 \vartheta
}+\mathrm{v}_{3}\hbar ke^{-i \vartheta }.
\end{equation}

Taking into account Eqs. (\ref{hamsec})-(\ref{f_oper}), the second
quantized Hamiltonian for the single-particle part can be expressed
as
\begin{equation}
\widehat{H}=\sum\limits_{\mathbf{k,}\sigma }\mathcal{E}_{
\mathbf{k}\sigma } \widehat{a}_{\mathbf{k}\sigma
}^{+}\widehat{a}_{\mathbf{k}\sigma } +e\mathbf{E}\left( t\right)
\sum\limits_{\mathbf{k,}\sigma }\sum\limits_{
\mathbf{k}^{\prime},\sigma ^{\prime }}\mathbf{D}_{\sigma
\sigma^{\prime }}\left( \mathbf{k,k}^{\prime }\right)
\widehat{a}_{\mathbf{k}\sigma }^{+}\widehat{a}_{\mathbf{k}^{\prime
}\sigma ^{\prime }},  \label{Ham}
\end{equation}

For the dipole matrix element for transition between the valence and
conduction bands

\begin{equation}
d_{\sigma \sigma ^{\prime }}\left( \mathbf{k,k}^{\prime }\right) =ie\mathbf{D%
}_{\sigma \sigma ^{\prime }}\left( \mathbf{k,k}^{\prime }\right) ,
\end{equation}
where
\begin{equation}
\mathbf{D}_{\sigma \sigma ^{\prime }}\left( \mathbf{k,k}^{\prime
}\right) =\psi _{\zeta ,\sigma }^{+}\left( \mathbf{k}\right) \psi
_{\zeta ,\sigma
^{\prime }}\left( \mathbf{k}^{\prime }\right) \frac{1}{S}\int \mathbf{r}e^{%
\frac{i}{\hbar }\left( \mathbf{k}^{\prime }-\mathbf{k}\right) \mathbf{r}}d%
\mathbf{r}.
\end{equation}
we obtained an analytical expression, which we do not explicitly
present here due to its long form. In contrast to the
nonrelativistic case, we found that the dipole matrix element for
the light interaction with the bilayer depends on the electron
momentum.

We use Heisenberg picture, where operator evolution is given by
the following equation%
\begin{equation}
i\hbar \frac{\partial \widehat{L}}{\partial t}=\left[ \widehat{L},\widehat{H}%
\right] ,  \label{Heis}
\end{equation}%
The single-particle density matrix in momentum space is defined as:%
\begin{equation}
\rho _{\sigma _{1}\sigma
_{2}}(\mathbf{k}_{1},\mathbf{k}_{2},t)=<\widehat{a}
_{\mathbf{k}_{2},\sigma_{2}}^{+}(t)\widehat{a}_{\mathbf{k}_{1},\sigma
_{1}}(t)>. \label{SPDM}
\end{equation}%

Using equations Eqs.(\ref{Ham})-(\ref{SPDM}) one can arrive at the
evolution equation for the single-particle density matrix for
different values of $\mathbf{k}$ and in external laser field:

\begin{equation}
i\hbar \frac{\partial \rho _{1-1}(\mathbf{k},t)}{\partial t}=2
\mathcal{E}_{ \mathbf{k}1} \rho _{1-1}(\mathbf{k} ,t)+E\left(
t\right) d\left( \mathbf{k}\right) \left[ \rho
_{11}(\mathbf{k},t)-\rho _{-1-1}(\mathbf{k},t)\right], \label{nleq1}
\end{equation}
\begin{equation}
i\hbar\frac{\partial \rho _{11}(\mathbf{k},t)}{\partial t}= [ \rho
_{1-1}(\mathbf{k},t)(E(t)d(\mathbf{k}))^{\ast}  - c.c.],
\label{nleq2}
\end{equation}

where the index ($\sigma =1$) is connected with the conduction band
and the index ($\sigma =-1$) is associated with the valence band,
$\rho _{11}$  and $\rho _{-1-1}$ are the populations in the
conduction and valence bands correspondingly, while $\rho _{1-1}$
describes the interband polarisation induced by the laser field.

\section{MANY-BODY CORRELATIONS IN BILAYER GRAPHENE} \label{sec3}

The Hamiltonian for bilayer graphene in the second quantization
formalism, in the presence  of electron-electron interaction, has
the form
\begin{equation}
{H} =\int {\Psi}^{+}H_{s}{\Psi}d\mathbf{r} \label{hamsec1}
+{H}_{Coul},
\end{equation}

\begin{eqnarray}
{H}_{Coul} =\int \int {\Psi}^{+}(\mathbf{r)}{\Psi}^{+}(\mathbf{r}^{\prime }%
\mathbf{)}V(|\mathbf{r}-\mathbf{r}^{\prime }|\mathbf{)}{\Psi}(\mathbf{r}%
^{\prime }\mathbf{)}{\Psi}(\mathbf{r)}d\mathbf{r}^{\prime
}d\mathbf{r}. \label{hamCoul1}
\end{eqnarray}
In Eq. (\ref{hamsec1}) the Hamiltonian ${H}_{s}~$ for bilayer
graphene in the presence of laser field  is given by
Eq.(\ref{hamsin}) and $\hat{H}_{Coul}$ is the Coulomb Hamiltonian.

Using the expansion of the fermionic field operator over the
annihilation and creation operators Eq.(\ref{f_oper}), we obtain the
expression for total Coulomb Hamiltonian, that consists of the four
terms: $ {H}_{Coul}={H}_{I}+{H}_{II}+{H}_{III}+ {H}_{IV}$.

Due to its complexity, we bring here the expression only for the
first term:

\begin{eqnarray*}
{H}_{I} &=&\frac{1}{2}\sum_{\mathbf{\mathbf{k}^{\prime }\mathbf{k}%
^{\prime \prime }q}}V_{2D}\left( \mathbf{q}\right) \{ \\
&&\widehat{c}_{\mathbf{k}^{\prime}-\mathbf{q}}^{+}\widehat{c}_{\mathbf{k''}+%
\mathbf{q}}^{+}\widehat{c}_{\mathbf{k}^{\prime \prime }}\widehat{c}_{\mathbf{%
k}^{\prime }}f^{+}(\mathbf{k}^{\prime }-\mathbf{q})f^{+}(%
\mathbf{k''}+\mathbf{q})f(\mathbf{k''})f(\mathbf{k}^{\prime }) \\
&&+\widehat{c}_{\mathbf{k}^{\prime }-\mathbf{q}}^{+}\widehat{c}_{\mathbf{k''}+%
\mathbf{q}}^{+}\widehat{c}_{\mathbf{k}^{\prime \prime }}\widehat{v}_{\mathbf{%
k}^{\prime }}f^{+}(\mathbf{k}^{\prime }-\mathbf{q})f^{+}(%
\mathbf{k''}+\mathbf{q})~f(\mathbf{k''})f_{v}(\mathbf{k}^{\prime }) \\
&&+\widehat{c}_{\mathbf{k}^{\prime}-\mathbf{q}}^{+}\widehat{c}_{\mathbf{k''}+%
\mathbf{q}}^{+}\widehat{v}_{\mathbf{k}^{\prime \prime }}\widehat{c}_{\mathbf{%
k}^{\prime }}f^{+}(\mathbf{k}^{\prime }-\mathbf{q})f^{+ }(%
\mathbf{k''}+\mathbf{q})f_{v}(\mathbf{k''})f(\mathbf{k}^{\prime }) \\
&&+\widehat{c}_{\mathbf{k}^{\prime }-\mathbf{q}}^{+}\widehat{c}_{\mathbf{k''}+%
\mathbf{q}}^{+}\widehat{v}_{\mathbf{k}^{\prime \prime }}\widehat{v}_{\mathbf{%
k}^{\prime }}f^{+}(\mathbf{k}^{\prime }-\mathbf{q})f^{+}(%
\mathbf{k''}+\mathbf{q})f_{v}(\mathbf{k''})f_{v}(\mathbf{k}^{\prime
})\} \label{Coul1}
\end{eqnarray*}

Here we define the creation operator $\widehat{a}_{\mathbf{k},\sigma
}^{+}$ in the conduction band as $\widehat{c}_{\mathbf{k}}^{+}(t)$,
and annihilation operator $\widehat{a}_{\mathbf{k},\sigma }$ in the
valence band as $\widehat{v}_{\mathbf{k}}(t)$ and $f(k)$ is the the
wave function Eq.(\ref{spinor}) in the specific $\textbf{k}$ point.

 As we see, the first Coulomb Hamiltonian
${H}_{I}$, by itself, consists of four terms: ${H}_{I}\
=~{H}_{I}^{1}+ {H}_{I}^{2}+{H}_{I}^{3}+{H}_{I}^{4}~$, and the total
${H}_{Coul}$ consists of 16 terms.

The second term of Coulomb Hamiltonian $H_{II}$ can be obtained from
$H_{I}$ by changing in the second column the operator related to the
conduction band $\widehat{c}_{\mathbf{k''+q}}^{+}$  by the valence
one $\widehat{v}_{\mathbf{k''+q}}^{+}$. These operators satisfy  the
anticommutation rules:
$[\widehat{c}_{\mathbf{k}}^{+},\widehat{c}_{\mathbf{k} ^{\prime
}}]_{+} =\delta _{ \mathbf{k,k^{\prime }}}\ $, $
[\widehat{c}_{\mathbf{k}}, \widehat{c}_{\mathbf{k} ^{\prime
}}]_{+}=0 $, $ [\widehat{c}_{\mathbf{k}}^{+},\widehat{v}_{\mathbf{k}
^{\prime }}]_{+}=0 $.

To take into account the contribution of the Coulomb interaction for
an operator evolution in Eq.(\ref{Heis}), we use now the total
Hamiltonian given by Eq.(\ref{hamsec1}).

The product of four field operators describes all many particle
correlations as trions, biexcitons, etc. To take into account only
excitonic effects, we apply the Hartree-Fock approximation to the
many particle system, i.e. we express four field operator averages
in Coulomb Hamiltonian as products of the polarization and
population, e.g. for $<\widehat{v}_{\mathbf{k}}^{+}
\widehat{c}_{\mathbf{k} ^{\prime }+\mathbf{q}}
\widehat{c}_{\mathbf{k}+\mathbf{q}} \widehat{c}_{\mathbf{k}^{\prime
}}>$ we have
 \begin{equation}
 <\widehat{v}_{\mathbf{k}}^{+} \widehat{c}_{\mathbf{k}
^{\prime }+\mathbf{q}} \widehat{c}_{\mathbf{k}+\mathbf{q}}
\widehat{c}_{\mathbf{k}^{\prime }}>=< \widehat{v}_{\mathbf{k}}^{+}
\widehat{c}_{\mathbf{k}^{\prime }}><
\widehat{c}_{\mathbf{k'}+\mathbf{q}}^{+}
\widehat{c}_{\mathbf{k}+\mathbf{q}}>\delta _{ \mathbf{k,k^{\prime
}}}\ - < \widehat{v}_{\mathbf{k}}^{+}
\widehat{c}_{\mathbf{k}+\mathbf{q}}><
\widehat{c}_{\mathbf{k'}+\mathbf{q}}^{+}
\widehat{c}_{\mathbf{k}^{\prime }}>\delta _{ \mathbf{k,k+q}}
\label{rpa}
\end{equation}
Using this approximation, we truncate the infinite Bogolubov chain,
and obtain closed set of equations which leads to exact exciton
energy. Using  Eqs. (\ref{Ham})-(\ref{rpa}) we arrive at the
evolution equations for the single-particle density matrix for
different values of $k_{x}, k_{y}$ and in the field of external
laser, namely
\begin{equation}
i\hbar \frac{\partial \rho _{1-1}(\mathbf{k},t)}{\partial
t}=\Sigma(\textbf{k})\rho _{1-1}(\mathbf{k}
,t)+\Lambda(\textbf{k})\left[2 \rho _{11}(\mathbf{k},t)-1\right],
\end{equation}
\begin{equation}
i\hbar \frac{\partial \rho _{11}(\mathbf{k},t)}{\partial t}=[ \rho
_{1-1}(\mathbf{k},t) \Lambda(\textbf{k})^{\star}-\rho _{-1
1}(\mathbf{k},t) \Lambda(\textbf{k})],
\end{equation}
where
\begin{equation}\Sigma(\textbf{k})=2\mathcal{E}_{\mathbf{k}1}+\sum\limits_{
\textbf{q}\neq0}V_{2D}\left( \mathbf{q}\right)
T(\mathcal{E}_{\mathbf{k}1}, U, \rho _{1-1}(\mathbf{k}), \rho
_{11}(\mathbf{k})),
\end{equation}

\begin{equation}
\Lambda(\textbf{k})=E(t) d(\mathbf{k})+\sum\limits_{
\textbf{q}\neq0}V_{2D}\left( \mathbf{q}\right)
P(\mathcal{E}_{\mathbf{k}1}, U, \rho _{1-1}(\mathbf{k}), \rho
_{11}(\mathbf{k})).
\end{equation}

We do not show here the obtained analytical expressions for the
functions $T$ and $P$ because of their complicated forms. Notice,
that for graphene systems the functions $T$ and $P$ depend on the
value of the opened gap as well as on the energy (Eq.
(\ref{energ})).

\section{EXCITONIC ABSORPTION IN GAPPED MONOLAYER GRAPHENE} \label{sec4}
The potential energy difference, that opens an energy gap $U$, can
be achieved in monolayer graphene by the inversion symmetry
breaking. Experimentally, it can be obtained by placing graphene
onto a substrate  in which the $A$ and $B$ atoms experience
different on-site energies. As shown in  Ref.~\cite{zhou}, graphene
grown epitaxially on $SiC$ has a band gap of about 0.2 eV, owing to
the graphene-substrate interaction.

Here we consider the interaction of a weak electromagnetic wave with
monolayer graphene in linear regime, when energy gap is opened. We
assume again that the laser pulse  propagates in the perpendicular
direction to graphene plane $(XY)$ and the electric field $E(t)$ of
pulse with linear polarization lies in the graphene plane. We
consider only low excitation regime with a small density of
electrons and holes and concentrate on the analysis of one
electron-hole pair effect. We use the same method, developed for
bilayer graphene and based on second quantized technique. The
difference is, that the Hamiltonian of gapped graphene monolayer, in
the vicinity of the $K$point, has now the form

\begin{equation}
\widehat{H}_{0}=\left(
\begin{array}{cc}
-\frac{U}{2} & \mathrm{v}_{F}\hbar\left( {k}_{x}-i{k}_{y}\right) \\
\mathrm{v} _{F} \hbar\left( {k}_{x}+i{k}_{y}\right) & \frac{U}{2}
\end{array}
\right), \label{hammon}
\end{equation}

The following calculations do not depend on $s$, only the spin
summation leads to an extra 2 factor in the final result for
macroscopic interband polarization, that is induced by an
electromagnetic wave. Using macroscopic interband polarization, we
compute the optical susceptibility from which we can get the
absorption coefficient.

 Now, we expand the fermionic field operator
(see Eq.(\ref{f_oper})) over the free wave functions  of gapped
monolayer:
\begin{equation}
\psi _{\sigma }(\mathbf{k}) =\sqrt{\frac{\mathcal{E}_{ \mathbf{k}
\sigma }+U/2}{2\mathcal{E}_{ \mathbf{k}\sigma }}}\left(
\begin{array}{c}
1 \\
\mathrm{v}_{F}\hbar ke^{i \vartheta }/(\mathcal{E}_{
\mathbf{k}\sigma }+U/2)
\end{array}
\right)\label{spinor1}
\end{equation}

\begin{equation}
\psi _{\sigma }(\mathbf{k})^{+} =\sqrt{\frac{\mathcal{E}_{
\mathbf{k} \sigma }+U/2}{2\mathcal{E}_{ \mathbf{k}\sigma }}}( 1,
\mathrm{v}_{F}\hbar ke^{-i \vartheta }/(\mathcal{E}_{
\mathbf{k}\sigma }+U/2) ), \label{spinor2}
\end{equation}

where $\mathrm{v}_{F}=\sqrt{3}\gamma_0a/2\hbar$.

The expression for energy spectrum of monolayer graphene is:

\begin{eqnarray}
\mathcal{E}_{ \mathbf{k}\sigma } &=&\sigma \sqrt{\frac{U^{2}}{4}+(
\mathrm{v}_{F}\hbar k)^{2}} \label{energ1}
\end{eqnarray}

For the dipole matrix element $d_{\sigma \sigma ^{\prime }}\left(
\mathbf{k}\right) $ for monolayer graphene we obtained the
expression:
\begin{equation}
d_{\sigma \sigma ^{\prime }}\left( \mathbf{k}\right)=\frac{e\hbar
\mathrm{v}_{F}}{2\mathcal{E}_{ \mathbf{k}1}^{2}k}(-\mathcal{E}_{
\mathbf{k}1}k_{y}+iU k_{x}/2)\label{dipmon}
\end{equation}

The diagonal elements of the density matrix Eq.(\ref{SPDM}) are
populations in the valence and conduction bands, correspondingly,
while non-diagonal term is the interband polarization:

\begin{equation}
P_{\mathbf{k}}(t)=<\widehat{v}
_{\mathbf{k}}^{+}(t)\widehat{c}_{\mathbf{k}}(t)>, \label{polariz}
\end{equation}
\begin{equation*}
P_{\mathbf{k}}^{\ast}(t)=<\widehat{c}
_{\mathbf{k}}^{+}(t)\widehat{v}_{\mathbf{k}}(t)>.
\end{equation*}

In order to find time dependent equations for the interband
polarization we use Heisenberg picture, where operator evolution,
e.g. for $\widehat{v} _{\mathbf{k}}^{+}(t)$  is given by the
following equation
\begin{equation}
i\hbar \frac{\partial \widehat{v} _{\mathbf{k}}^{+}(t)}{\partial
t}=\left[ \widehat{v} _{\mathbf{k}}^{+}(t),H \right] ,
\label{Heis1}
\end{equation}%
where the Hamiltonian $H$ is given by Eq. (\ref{hamsec1}).

 To take into account only excitonic effects, we again apply
the Hartree-Fock approximation to the many particle system.
 In linear regime we assume that the population in the conduction band is almost zero,
 while the population in the valence band is unity. In this regime we have only equation for interband polarization,
 and that takes into account the Coulomb interaction.
Using Eqs. (\ref{hamsec1},\ref{hamCoul1}) and Eqs.
(\ref{hammon})-(\ref{Heis1}) we obtain the following two equations
for the real and imaginary parts of the polarization
$P_{\mathbf{k}}(t)=P'_{\mathbf{k}}+iP''_{\mathbf{k}}=P'_{k
\vartheta}+iP''_{k \vartheta}$

\begin{equation}
\hbar \frac{P'_{\mathbf{k}}(t)}{\partial
t}=2\mathcal{E}_{\mathbf{k}1}P''_{\mathbf{k}}(t)-Im[E(t)
d(\mathbf{k})]+\sum\limits_{ \textbf{q}\neq0}V_{2D}\left(
\mathbf{q}\right)[(U
sin(\vartheta-\vartheta')/2\mathcal{E}_{\mathbf{k}1})P'_{\mathbf{k}}
+cos(\vartheta-\vartheta')P''_{\mathbf{k}}]
\end{equation}
\begin{equation}
\hbar \frac{P''_{\mathbf{k}}(t)}{\partial
t}=-2\mathcal{E}_{\mathbf{k}1}P'_{\mathbf{k}}(t)+Re[E(t)
d(\mathbf{k})]+\sum\limits_{ \textbf{q}\neq0}V_{2D}\left(
\mathbf{q}\right)[\frac{(\gamma_{0}^{2}kk'+U^{2}cos(\vartheta-\vartheta')}{4\mathcal{E}_{\mathbf{k}1}\mathcal{E}_{\mathbf{k'}1}}
)P'_{\mathbf{k}}-\frac{U
sin(\vartheta-\vartheta')}{2\mathcal{E}_{\mathbf{k}1}}P''_{\mathbf{k}}],
\end{equation}

where  $V_{2D}( \mathbf{q})$  is the Fourier transform of
two-dimensional Coulomb interaction,  $d(\mathbf{k})$ is defined by
Eq.(\ref{dipmon}), and the electric field of laser pulse has the
following form: $E(t)=exp({-i \omega t }) exp[({-t/\tau})^{2}]$ with
$\tau\gg T=2\pi/\omega$. We solve the obtained integro-differential
equations numerically using Runge-Kutta  method with time step
$\Delta t \ll T $. As a result, we obtain the interband polarization
as a function of time, and define the macroscopic polarization that
includes contributions from different values  of $\textbf{k}$ :
$P(t)=\sum\limits_{ \textbf{k}}P_{\mathbf{k}}(t)
d^{\ast}(\mathbf{k})+c.c.$

Using Fourier transform of macroscopic interband polarization, we
compute the optical susceptibility, and get the absorption
coefficient as the imaginary part of the optical susceptibility.

\section{RESULTS AND DISCUSSION} \label{sec5}
\begin{figure}
\centering
\includegraphics*[width=15cm]{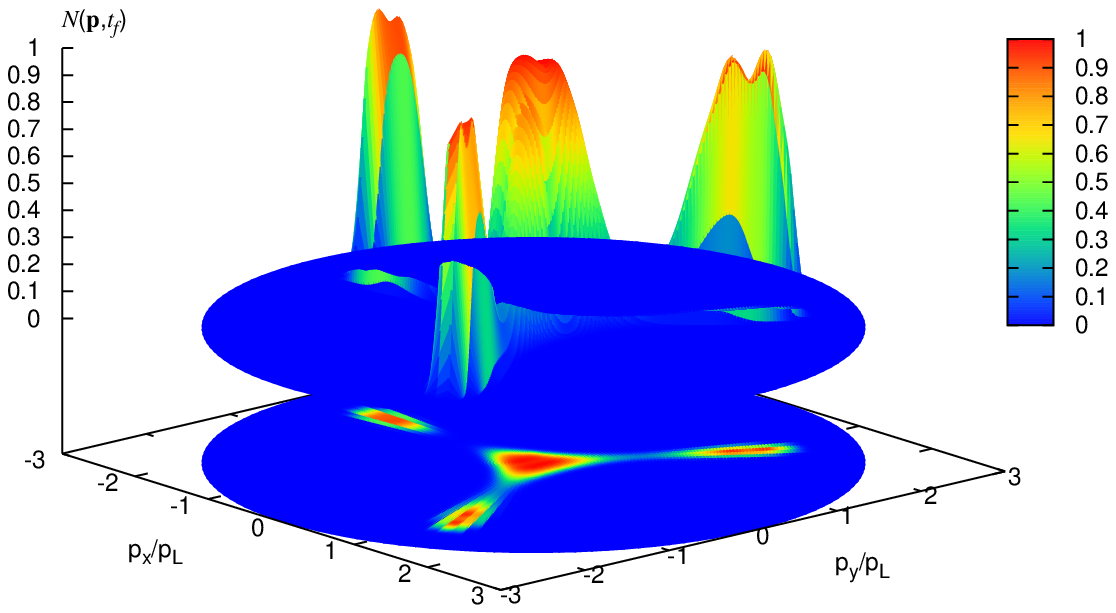}\caption{
Particle distribution function $N_{c}(p)$ (in arbitrary units) after
the interaction with the pulse having $\hbar\omega_0=8 E_L$ ( when
$\hbar\omega_0\approx U_{fin}$) as a function of scaled
dimensionless momentum components.}\label{2Dband8E}.
\end{figure}

\begin{figure}
\centering
\includegraphics*[width=15cm]{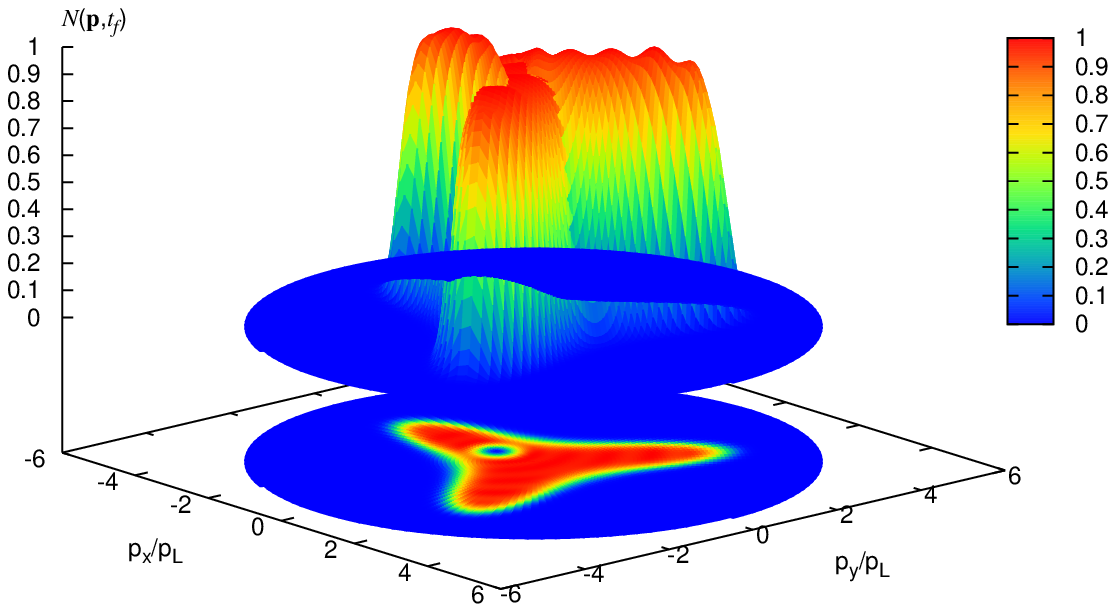}\caption{
Particle distribution function $N_{c}(p)$ (in arbitrary units) after
the interaction with the pulse with $\hbar \omega_0= 50 E_L$ ( when
$\hbar\omega_0\approx U_{fin}$) as a function of scaled
dimensionless momentum components.}\label{2Dband50E}
\end{figure}

We solve numerically differential equations
Eqs.(\ref{nleq1},\ref{nleq2}) for single-particle density matrix
using Runge-Kutta method. Fig.~\ref{2Dband8E} shows 2D plot for the
electron distribution in the conduction band $\rho_{11}=N_c$ after
the interaction with the laser pulse with $\hbar \omega_0= 8 E_L$ as
a function of dimensionless momentum components (in units of the
Lifshitz energy $E_L=m \mathrm{v_{L}}^{2}/2=1 meV$ and momentum
$p_L=m \mathrm{v_{L}}$). Fig.~\ref{2Dband50E} shows the electron
distribution function $\rho_{11}=N_c$ in the conduction band for
$\hbar \omega_0= 50 E_L$.

We see that for larger value of the frequency, the light grey
triangle that corresponds to higher values of $N_c$, becomes larger,
and consequently a larger number of electrons is transferred to the
conduction band. It is connected with the fact, that for larger
value of the frequency, the resonant value of the gap is larger: for
such values of the gap, the bands in bilayer near the $K$ point have
stronger flatness. Estimations show that the density of electrons in
the triangle is about $10^{11}cm^{-2}$. The high density of excitons
can lead to Bose condensation phenomenon.

In contrast with this, for small values of the gap the electron
transitions from the valence to the conduction band take place only
from separate regions of momentum space (see Fig.~\ref{2Dband8E}).
This is  due to the fact  that for low energies, close to the
Lifshitz energy $E_L$, the isoenergetic line is dropped into four
separate pockets~\cite{castrorev, avetis}. There is one central part
with $p=0$ and three "leg" pockets with $p=p_{0}$, so the
isoenergetic line is stretched in three directions (for valley $K$
the line is deformed along the directions $\varphi_{0}=0,2/3\pi$ and
$4/3\pi$). For small photon energies, i.e. in the case, when $\hbar
\omega_0= 8 E_L$, when the resonance takes place, the influence of
this effect is more observable (see  small light regions on the
triangle in Fig.~\ref{2Dband8E}).

\begin{figure}
\centering
\includegraphics*[width=12cm]{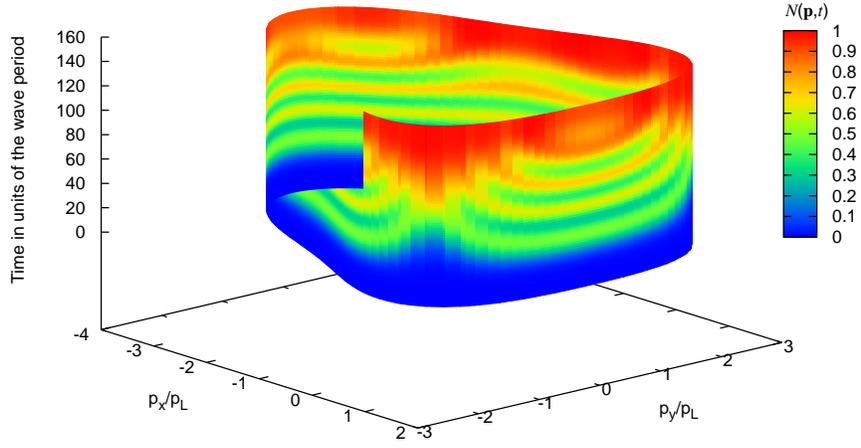}\caption{
The evolution of the particle distribution function
$N_{c}(p)$ (in arbitrary units) during the interaction with the pulse with $%
\hbar \omega_0= 32 E_L$, when the energy gap of bilayer graphene
reaches its maximal final value $U_{fin}=28 E_L$}
 \label{inversev}
\end{figure}

The  obtained 2D plot for the evolution of particle distribution
function in the conduction band $\rho
_{cc}(\mathbf{p},\mathbf{p},t)=<\widehat{c}
_{\mathbf{k}}^{+}(t)\widehat{c}_{\mathbf{k}}(t)>=N_{c}(p)$ with time
is shown in Fig.~\ref{inversev} for the pulse having $\hbar
\omega_0= 32 E_L$ with $\omega_0$ its frequency (in units of $E_L$
and $p_L$). For this case, during the interaction time $t_{f}=100T$,
the energy gap of bilayer graphene reaches its maximal final value
$U_{fin}=28 E_L$. We see that for $\hbar \omega_0> U_{fin}$ the
electrons transfer to the conduction band in the region which is
higher than the bottom of the conduction band. Fig.~\ref{inversev}
reflects the band structure of gapped bilayer graphene with trigonal
shape of energy bands which takes place due to $\gamma_{3}$
interaction~\cite{avetis}. As shown in Fig.~\ref{inversev}, in the
beginning of the interaction, the population of electrons  in the
conduction band is negligible (that corresponds to dark contour in
Fig.~\ref{inversev}) while in the end of the interaction we observe
the full inversion of the population of electrons between the
valence and the conduction bands (light contour in
Fig.~\ref{inversev}).

\begin{figure}
\centering
\includegraphics*[width=12cm]{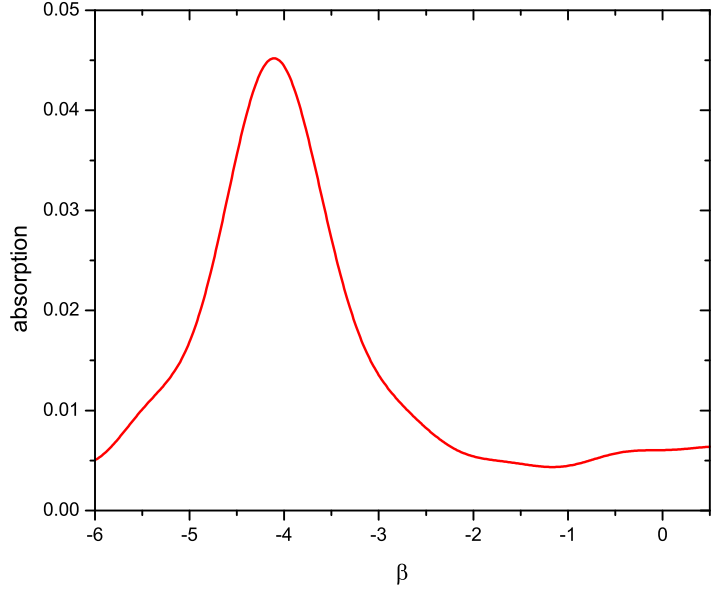}\caption{
Excitonic absorption in monolayer graphene as a function of the
detuning $\beta=(U-\hbar \omega)/E_R$ for the value of effective
fine structure $\alpha=0.175$} \label{absorp}
\end{figure}

Fig.~\ref{absorp} shows excitonic absorption spectrum of gated
monolayer graphene with the energy gap $500meV$ as a function of
detuning $\beta=(U-\hbar \omega)/E_R$ where $E_{R}=\mu
e^{4}/2\hbar^{2}\chi^{2}$ is the effective Rydberg energy. We
introduce the reduced effective electron-hole mass by the expression
$\mu=U/4\mathrm{v_F}^{2}$; the mass is simply proportional to the
band gap as in  a simple two-band model  proposed in~\cite{peders}.
In contrast to the standard semiconductor case, in graphene one
deals with fine structure constant  , and the appearance of bound
states strongly depends on the value of $ \alpha=\chi
e^{2}/\hbar\mathrm{v_{F}}$ (the choice of dielectric constant
strongly affects this). For $\alpha=0.175$  (for the value of
dielectric constant $\chi=12.5$ ) we obtained the maximum of the
excitonic peak at $4.12E_{R}$ (see Fig.~\ref{absorp}).

This result is in a good agreement with the exact analytical
solution for relativistic 2D hydrogen atom ~\cite{guo}. The
expression for the ground state energy in Ref.~\cite{guo} has the
form $ E_{c}=m_{e}c^{2}[1+4\alpha^{2}/(1-4 \alpha^{2})]^{-1/2}$
which gives $E\approx4.13 E_{R}$  for the value of $\alpha=0.175$ .

We found that the width of the peak in Fig.~\ref{absorp} is much
larger than in nonrelativistic case.

In a future paper, and on the basis of the above developed method,
we plan to show our investigation of excitonic absorption in gaped
bilayer graphene.

\section{Conclusions} \label{sec6}

In the present paper the microscopic theory of a strong
electromagnetic radiation interaction with multilayer graphene
systems is developed. We consider one-photon resonant interaction of
a laser field with bilayer graphene when an energy gap $U$  is
opened due to external gates. We found that changing the energy gap
linearly with time, the electron population is transferred from the
top of valence band to the bottom of conduction one after the time
$t_{1}$, when the gap comes into resonance with the electromagnetic
field.

It is well known that a frequency-chirped laser pulse may produce
full inversion of the populations between the ground and excited
states in the two-level atom in the adiabatic following
approximation [25, 26]. The population transfer in graphene systems
can also be achieved by using traditional frequency chirped
electromagnetic pulses, but the suggested method (i.e. adiabatic
change of the energy gap) is more convenient for graphene systems,
since there are technical difficulties with terahertz radiation
manipulation.

In this sense, graphene devices can in turn be used for infrared and
terahertz radiation detection and frequency conversion.

We also found that due to relative flatness of the bottom (top) of
conduction (valence) band in multilayer graphene systems in the
presence of perpendicular electric field, the density of coherently
created particle-hole pairs becomes quite large, which can make
Bose-Einstein condensation of electron-hole pairs possible.

 We consider also excitonic states in monolayer graphene with opened
energy gap. To take into account the Coulomb interaction, we use
Hartree-Fock approximation that leads to a closed set of equations
for the single-particle density matrix, which in turn produce our
final results for the excitonic absorption in this case. A broader
investigation of excitonic absorption for gated bilayer graphene is
reserved for a future article currently under preparation.

\begin{acknowledgments}
The Authors thank Dr. G.F. Mkrtchian and Prof. F.M. Peeters for
fruitful discussions.
\end{acknowledgments}

\clearpage

\clearpage

\clearpage

\clearpage

\end{document}